\documentclass{sf2a-conf2021}
\usepackage{graphicx}
\usepackage{hyperref}
\usepackage[]{natbib}  
\usepackage{epstopdf}

\def\BibTeX{{\rm B\kern-.05em{\sc i\kern-.025em b}\kern-.08em
    T\kern-.1667em\lower.7ex\hbox{E}\kern-.125emX}}
\bibpunct{(}{)}{;}{a}{}{,}  %%%%%%%%%%%%%  A&A bibliography style
\begin{document}
\TitreGlobal{SF2A 2021}

%%-----------------------------------------------------------------
%%      the top matter
\title{Science with SKA}
\runningtitle{Science with SKA}

\author{F. Combes}\address{Observatoire de Paris, LERMA, Coll\`ege de France,
CNRS, PSL University, Sorbonne University, 75014, Paris, France}

%% Keep this line, even if the page will be settled afterwards.
\setcounter{page}{237}

%%-----------------------------------------------------------------

\maketitle

%%-----------------------------------------------------------------
%%        The abstract
%% 
%%  Warning!  within the abstract:
%%  - do not use macros. 
%%  - do not use commands like: \cite, \citet, \citep ... etc.

\begin{abstract}
Highlights are presented about the science to be done with SKA. as well as state of the art 
science already done today with its precursors (MeerKAT, ASKAP) and pathfinders 
(LOFAR, NenuFAR), with accent on the expected breakthroughs.
\end{abstract}

%% Insert the keywords (to appear in the ADS indexing)
%% Keywords must be separated by a comma

%%-----------------------------------------------------------------

\section{The key scientific projects of SKA}
%%---------------------
Some of the main puzzles of cosmology are the nature of dark matter 
and dark energy, representing in total 95\% of the content of the Universe. 
 The dark energy is presently compatible within the uncertainties with
 a cosmological constant, but it is paramount to determine with greater
 precision whether its evolution with time is dynamic, and could be due
 to a fifth element, a quintessence, and new physics. The tools for making
 this diagnostic are the same as for many other dark energy probes, either 
 from ithe ground or in space, with Euclid: the BAO (Baryon Acoustic 
 Oscillations), playing the role of a standard ruler, measuring the expansion
 at differnt redshifts, the Weak Lensing (WL), or Redshift Space Distorsions
 (RSD), measuring the density and amplitude of large-scale structures to
 constrain the evolution of $\Omega$ and $\Lambda$.  These tools will be 
 exploited with optical tracers, and the novelty of SKA is to use radio 
 tracers, and the HI-21cm line to identify galaxies. These tracers have
 different biases than the optical ones, and both studies 
 are very complementary.
 Optically, the massive galaxies are early-type gathered in galaxy clusters,
 while the HI-rich galaxies are late-type in the field.
 
 Another key project is to explore the Epoch of Reionization (EoR), likely
 to extend from z=20 to 6. If it is already possible to have some clues
 with present searches of galaxies and quasars at z$>$6, the inter-galactic
 medium will be uniquely explored by SKA, with the redshifted 21cm-HI line,
 as it is now with pathfinders and precursors. The large galaxy surveys made
 over the whole available sky, due to the wide field of view, will serve to
 determine uniquely the large-scale structures, and the galaxy formation and
 evolution.

 Pulsars will be discovered in a huge number with SKA, exploring the
 whole Milky Way, while presently they are confined in the solar neighborhood.
 Milli-second pulsars are extremely precise clocks, which can be used to
 detect very long wavelength gravitational waves. 
Strong-gravity will be explored with pulsars and black holes

Cosmic magnetism is another key project, and in particular the formation
of primordial magnetic fields will be tackled. Finally, the search for the
origin of life, the mapping of the protoplanetary disks, and the search
for pre-biotic molecules, will be carried out in synergy and complementarity
with ALMA at higher frequencies.

All key projects with SKA have been developed in many whitepapers and 
conferences, \citep[e.g.,][]{Carilli2004,Carilli2015}.

\section{Cosmology and galaxies}
%%-------------------------

\subsection{Dark sector and new physics}

 The state of the art constraints on the dark energy and
 the dark matter are obtained by combining all available data,
 from the SNIa standard candels, i.e. the 
Pantheon sample of 1048 SNIa between redshifts 0.01 $<$ z $<$ 2.3
 \citep{Scolnic2018}, and the 207 SN sample from DES-3yr
\citep{Abbott2019}, with the BAO results from SDSS \citep{Alam2021}, and the 
CMB data \citep{Planck2020}.
The equation of state of the dark energy can be written as the
pressure proportional to the density P =  w $\rho$, with w negative,
and the variation of w(a)= w$_0$ + wa (1-a), where a is the characteristic
radius of the Universe,  a=1/(1+z).
Since SNIa are difficult to observe at z$>$1, \citet{Inserra2021}
propose to use and calibrate superluminous supernovae (SLSNe), which will allow
to go farther and faster. The first results are promising, putting constraints
in the  w$_0$-wa diagram.  With the enhanced precision acquired in the
recent years, some tension has grown between observations and the standard
$\Lambda$CDM model, suggesting the necessity of new physics
\citep{Smith2020}. In particular the main tension occurs between the Hubble 
constant H$_0$=  73.48$\pm$1.66 km/s/Mpc
measured locally with Cepheids or other indicators
\citep{Riess2018}, and the Planck determination of 67.4$\pm$ 0.5 km/s/Mpc. 
The discrepancy reaches 3.7$\sigma$.
In radioastronomy, powerful masers (H$_2$O, OH..) allow to observe in VLBI the 
center of external galaxies, and their rotating circum-nuclear disk;
measuring the velocities through the Doppler effect and  monitoring  
through VLBI the gradient of maser poition with velocity,
 results in a precise distance indicator, as shown
beautifully by the prototypical example of NGC~4258 \citep{Greenhill1995}.
SKA will measure many more masers around AGN at various redshifts,
and can give a complementary approach to the problem.  Already
\citet{Pesce2020} with megamasers confirm H$_0$ = 73.9$\pm$3 km/s/Mpc.

The most recent BAO and RSD results, including 147 000 quasars
\citep{Ata2018}, and Ly$\alpha$ absorption surveys \citep{Bautista2017},
are compatible with a flat $\Lambda$CDM cosmology, and 
perfectly compatible with the Planck cosmological parameters
\citep{Bautista2021}.

SKA-1 surveys of galaxies in the HI-21cm lines will
be complementary and competitive with the optical ones from the ground
and in space (Euclid). Surveys where galaxies are detected individually will
be most useful for galaxy formation and evolution, they will detect
4 million galaxies up to z=0.2 in the all-sky survey, 2 million
galaxies up to z=0.6 in the wide field, and 0.4 million in the deep field
survey up to z=0.8 (of 50 square degrees area).
For cosmology purposes, HI intensity mapping over 30 000 
square degrees, and covering redshifts up to 3 will be 
more competitive \citep{Maartens2015}. Weak lensing in radio surveys up 
to redshift z=6 will consider a billion objects. One of the strong 
advantages of SKA-1 
is the much larger volumes sampled, with respect to all other probes
(Euclid, DESI, BOSS, Nancy-Grace-Roman...).
The second phase SKA-2 will surpass all.

%% figures side by side Fig 1 
\begin{figure}[ht!]
 \centering
 \includegraphics[width=0.8\textwidth,clip]{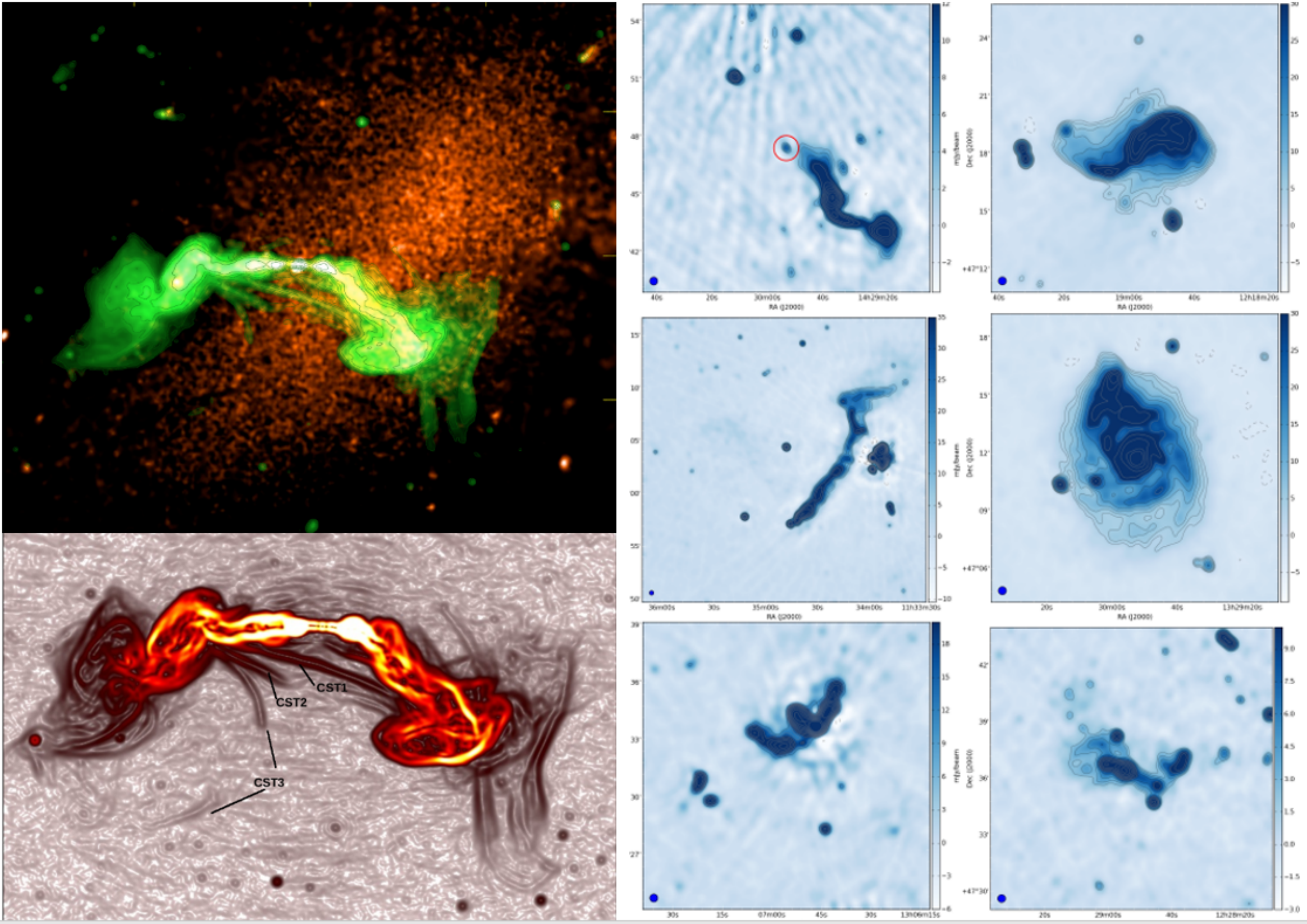}      
  \caption{{\bf Left:} Radio continuum emission from ESO 137-006 detected by MeerKAT at 1030 MHz.  Three collimated synchrotron threads (CSTs) between the 
radio lobes are indicated. 
The Sobel filter has been applied to the image, to better show these features
	\citep{Ramatsoku2020}. The upper panel shows
	that the galaxy is entering the Norma cluster, and its X-ray
	gas atmosphere (in red)..
	{\bf Right:} Some radio images of nearby galaxies from the 
LOFAR Two-metre Sky Survey (LoTSS) \citep{Shimwell2019}.}
  \label{fig1}
\end{figure}

\subsection{Continuum and HI surveys}

Simulations have been performed of wide-field images of the 
radio-continuum sky with SKA, detecting both the very numerous star-forming 
galaxies, with synchrtron emission coming from supernovae, and the 
stronger but less numerous radio AGN, of FRI and FRII types, the latter
even less numerous but stronger \citep{Jackson2004}. The AGN radio jets can 
be used easily as standard rods, constraining the cosmological parameters,
by themsleves, and also through weak lensing.

For the all-sky survey at 1.4 GHz, in 2yrs of integration,
SKA1 will achieve 3 $\mu$Jy rms, and detect $\sim$4 
galaxies per arcmin$^2$ (at more than 10$\sigma$),
\citep{Jarvis2015}. The survey will be made with an
excellent quality circular Gaussian beam from about 0.6 to 100'',
With almost uniform sky coverage of 3$\pi$ str.
This will provide a total of 0.5  billion radio sources, 
yielding weak  lensing and Integrated Sachs Wolfe (WL, ISW) diagnostics.
For the wide-field (5000 deg$^2$), with 2 $\mu$Jy rms $\sim$6 galaxies 
per arcmin$^2$  are expected (at more than 10$\sigma$).
For the deep-field (50deg$^2$) with 0.1 $\mu$Jy rms, $\sim$20 galaxies
per arcmin$^2$ will be detected,  at more than 10$\sigma$.

Figure \ref{fig1} shows some examples of radio images from the
LOFAR Two-metre Sky Survey (LoTSS), and also how the precursor MeerKAT
has discovered new features in typical radio-jets: collimated synchrotron 
threads, linking the radio lobes from the sides, in parallel to the radio jets,
in ESO 137-006. This galaxy is moving inside the wind of the intra-cluster gas,
entering the Norma cluster. The radio lobes are distorted and bent, and the 
threads look like relics of the previous radio jets, in previous episodes 
of ejection.

In HI-21cm surveys, SKA-1 will allow the imaging of substantial number 
of high-redshift galaxies for the first time \citep{Staveley2015}.
While the present instruments are restricted to detect HI in individual 
galaxies only to the local Universe up to z=0.1, the very deep survey
will permit the detection of galaxies at z=2, and even higher for SKA-2.
 A glance of what intensities could be detected is given by recent
 stacking to detect only "globally" some remote galaxies.
 With GMRT deep (117h) field,  \citet{Bera2019} have stacked 445 blue galaxies
 between 0.2 $<$ z $<$ 0.4, and obtained a 
detection at 7$\sigma$ of  M(HI) = 5 10 $^9$ M$_\odot$. 
Stacking the continuum to derive the star formation rate, they
derive a depletion time  of $\sim$ 9 Gyrs.
From GAMA survey,  imaged on DINGO-VLA,
\citet{Chen2021} have stacked HI cubelets on a sample of 3622 galaxies,
and obtained a clear detection, with FWHM of 60km/s.
 
\section{Reionization}
%%-------------------------

Intensity mapping is the only technique able to determine
the global quantities searched for in the EoR.
Continuum foregrounds are typically 1000 times brighter than the 
expected cosmological signal. The instrumental responses to bright foregrounds 
with extended and multiple sidelobes, forming a sea of confused
signals, depending on their location on the field of view, are a challenge
to understand and subtract away \citep{Santos2015}.
The foregrounds to be eliminated produce a perturbing signal,
which is not necessarily spectrally smooth
\citep{Switzer2015}. 

The LOFAR key project on EoR has observed more than 1000hours
on selected clean fields, with the least possible foreground emission.
However, even if the nominal sensitivity is reached to detect easily
the expected signal, the confusion by foregrounds has prevented to
draw any conclusion. Controlling the calibration, and cleaning for the 
sidelobes down to the low intensity level required is a long process.
While in 2017, only 0.5\% of data were understood and used \citep{Patil2017},
 recently up to 5\% of data has been understood and cleaned, 
 resulting in an upper limit of the EoR signal two orders of magnitude above
 the expected signal.
There still remains noise that could be due to residual emission 
from foreground sources or diffuse emission far away from the phase centre, 
polarization leakage, chromatic calibration errors, ionosphere, or 
low-level radiofrequency interference \citep{Mertens2020}

\section{Pulsars, Cosmic magnetism}
%%-------------------------
 
\subsection{Pulsars and gravitational waves}

The large number of pulsars to be discovered by SKA, in combination 
with its exceptional timing precision, will revolutionize the field of 
pulsar astrophysics. SKA will provide a complete census of pulsars in 
both the Galaxy and in Galactic globular clusters 
\citep{Cordes2004}. In the Milky Way, about 30 000 pulsars should
be present, and 10 000 milli-second ones. May be 20 000 pulsars will
be detectable in the whole Galaxy (while today we know only pulsars in the 
solar neighborhood).
Pulsars and compact objects  will allow unique tests of the
strong field limit of relativistic gravity and the equation of state at extreme densities. 

Through monitoring these pulsars, which are extremely precise clocks,
(up to 10$^{-15}$ in relative), gravitational waves of long wavelengths, 
of the order of light-yrs, could be detected, and pulsars are the only way.
The hope is to detect the merger of super-massive black holes, and also
the primordial waves, signature of inflation
\citep{Janssen2015}. A first preliminary detection
has been claimed with present telescopes \citep{Arzoumanian2020}.

\subsection{Fast Radio Bursts}

Since their discovery by \citet{Lorimer2007}, our knowledge of
Fast Radio Bursts (FRB) have grown considerably: they have been detected in 
large numbers, especially with the wide-field instrument CHIME:
540 are known, and the occurence has been estimated at 800 per day on the
whole sky \citep{CHIME-FRB2019}.
With SKA-MID, it will be possible to detect 100 FRB/yr,
with precise localisation \citep{Keane2018}. 
The nature of the FRB phenomenon is not yet clarified, although 
many repeaters have been detected, and one FRB has been associated 
with a Galactic magnetar, SGR 1935+2154 \citep{Bochenek2020}.
Due to their surface magnetic fields larger than 10$^{14}$ gauss, 
magnetars are the source of high-energy phenomena, where their magnetic
energy decays. A florilege of theories have been proposed to explain
the phenomena \citep{Platts2019}.

%%  figure 2
\begin{figure}[ht!]
 \centering
 \includegraphics[width=0.8\textwidth,clip]{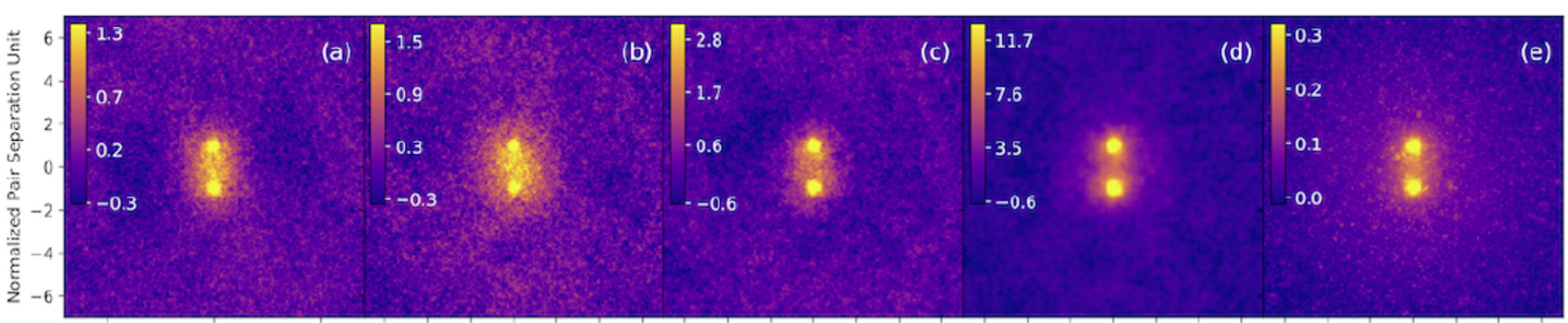}      
  \caption{Stacking radio and X-ray maps, around physically nearby pairs
of luminous red galaxies (LRG).
From left to right the columns are GLEAM 154 MHz,
 118 MHz,  88 MHz, OVRO-LWA 73 MHz, and ROSAT combined band 0.1 -2.4 keV. 
The colour bars all have
units of temperature in K, except the ROSAT maps which are in 
counts per second per arcmin$^2$ x10$^4$.
From \citet{Vernstrom2021}.}
  \label{fig2}
\end{figure}

\subsection{Magnetic Fields}

Polarisation of radio emission, and Faraday rotation have been used
intensively to determine the intensity and orientation of the magnetic
field in spiral galaxies, at various depths according
to the various wavelengths: either from the halo, or the disk,
and different distributions have been obtained with respect
to the spiral arms \citep{Kierdorf2020}. Turbulence due to star formation
in spiral arms paradoxically reduces alignment, and frequent field
reversals in the vertical direction contribute to distortions
that are not yet well understood. LOFAR has been used in combination with VLA
to determine the spectrum of the emission, separate thermal and
non-thermal components, the magnetic field strength and the cosmic ray
electron losses \citep{Mulcahy2018}.

All-sky survey of Faraday rotation will measure inter-galactic magnetic field, as 
well as inside galaxies. The mechanisms to generate the field are not yet
settled, from inflation, phase transitions in the early Universe, and 
batteries to amplify the seeds. Normally the field is  frozen in the ionized gas, but
should dilute away in the expansion. When structures collapse, the field is amplified
again \citep{Johnston2015}.

Searches have been done in diffuse filaments connecting clusters,
at the cosmic web  (15Mpc) scale, combining 
X-ray hot gas with eRosita with radio data from
ASKAP/EMU Early Science \citep{Reiprich2021}.
Missing baryons are searched for by studying
the warm-hot gas in cluster outskirts and filaments. 
The bridge between two clusters is detected; it may contain known galaxy
groupis, but not accounting for all the emission
There are several clumps of warm gas falling into the clusters,
compatible to what is observed in simulations.

LOFAR has also detected synchrotron emission in filaments 
between merging galaxies, with possible shocks re-accelerating
the electrons \citep{Govoni2019, Botteon2020}, but these were
only short scales. Now with GLEAM (the MWA survey), it is possible 
to search for longer filaments (see Figure \ref{fig2}).
These are traced by Luminous red galaxies (LRGs), which are
massive early-types  residing in the center of galaxies clusters or groups. 
The first large-scale filament detection, has revealed a magnetic
field of 30-60 nG, of intensity higher than previously believed,
with electrons subject to more  efficient shock acceleration
\citep{Vernstrom2021}.

%% figure 3
\begin{figure}[ht!]
 \centering
 \includegraphics[width=0.8\textwidth,clip]{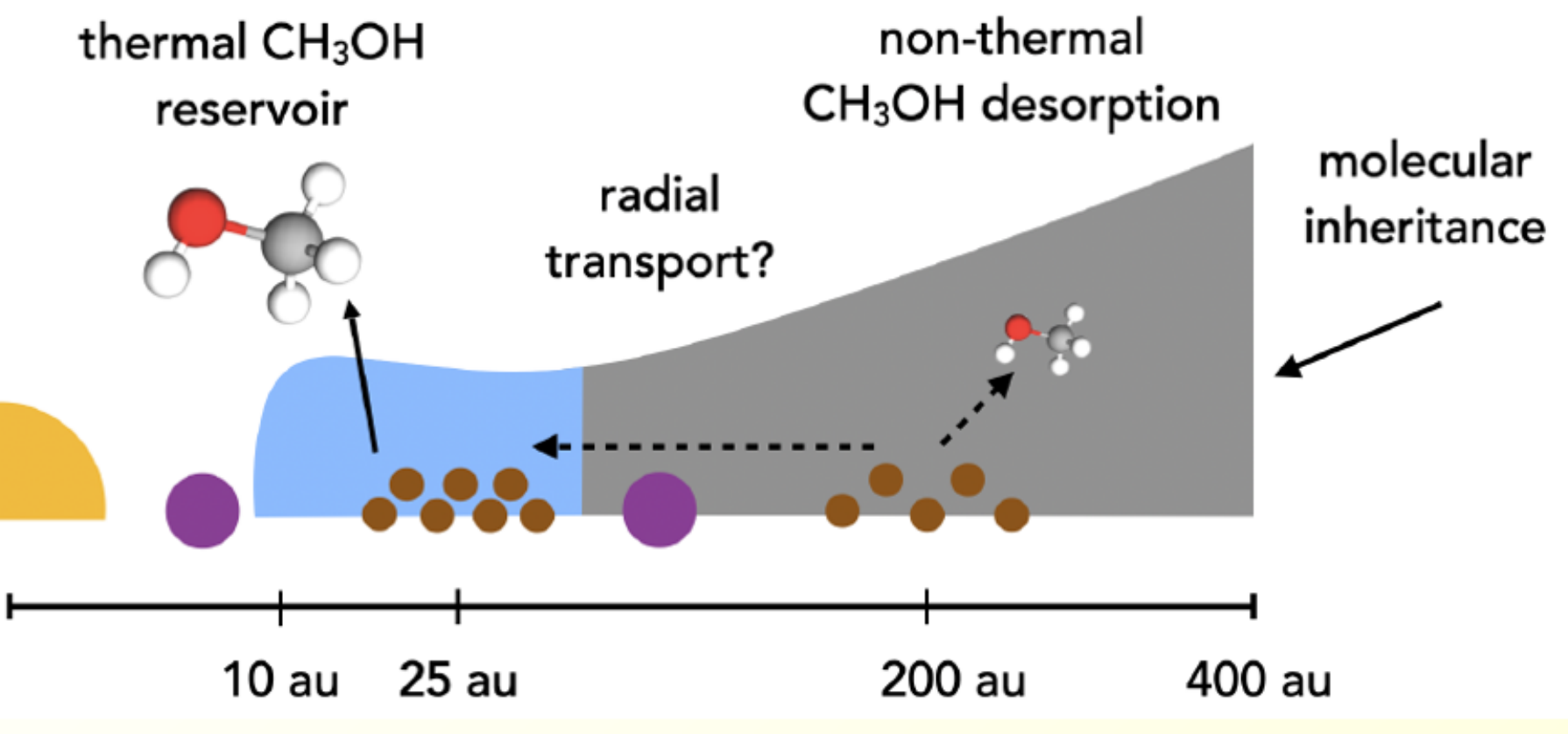}      
\caption{Cartoon of a protoplanetary disk (HD 100546) showing the regions where
CH$_3$OH is detected and the different physical and 
chemical mechanisms proposed \citep{Booth2021}.} 
  \label{fig3}
\end{figure}

\section{Craddle for life}
%%-------------------------

ALMA has made a breakthrough in the domain of the formation of planets
and protoplanetary disks, in imaging with superb resolution resonant
rings and gaps \citep{Andrews2020}.
 Disks are formed first with gas and small-size dust grains, 
 the latter agglomerating
 progressively in mm- and cm-sized grains before becoming planetesimals.
 These grains emit at longer and longer wavelengths, and SKA-1 will
 be the prefered instrument to detect cm to m-sized dust. At high
 resolution with 40mas beam, the nearest systems will be mapped
 with 4AU resolution, sufficient to determine the snow line
\citep{Hoare2015}.
Large exoplanets, of Jupiter-size, could be detected with their
magnetic fields \citep{Zarka2018}.
In synergy with ALMA, the detection of complex organic molecules (COM)
could be carried out, such as the methanol CH$_3$OH, with deuterated
species CH$_2$DOH, methanethiol CH$_3$SH, formamide NH$_2$CHO, and 
heavier pre-biotic molecules in Band 5, such as amino acids and sugars. In 
1000h SKA1-mid could detect clearly $\alpha$-alanine, with a hundred
of lines, for a column density of 10$^{13}$ cm$^{-2}$  
\citep{Hoare2015}.

Methanol has been detected in protoplanetary disks,
and is thought to come from hydrogenation of CO on icy grains
(cf Figure \ref{fig3}). Complex organic molecules have
now been detected, which are key to form amino-acids, and pre-biotic 
molecules \citep{Booth2021}. These COM cannot form in situ,
but must come from ices formed previously in dark interstellar clouds.

\section{Conclusions}
%%--------------------

SKA-1 will help to tackle the main puzzles of 
cosmology: the nature of dark matter and dark energy,
by using the common tools (BAO, WL, RSD) with
high precision, and with tracers with different
biases than optical surveys (radio continuum, HI in galaxies).
With extragalactic masers, it will be possible to bring a
complementary constraint to the H$_0$ problem, suggesting new
physics.

With redshifted HI-21cm line, SKA will have a unique contribution
to the Epoch of Reionization, and the birth of
the first galaxies. With the timing of millisecond pulsars, 
SKA will make a breakthrough in probing strong ggavity,
and detecting gravitational waves in the very low frequency regime
(nanoHz), to search for primordial waves.

Our knowledge of magnetic genesis will considerably improve. SKA will work in synergy with ALMA to determine the physics
of protoplanetary disks, and detect pre-biotic molecules.
 
All these key projects have begun to be tackled at very low
frequency with the NenuFAR pathfinder in Nan\c{c}ay, where the first
large programs are 
ES1: Cosmic Dawn;
ES2: Exoplanets \& Stars;
ES3: Pulsars;
ES4: Transients;
ES5: Fast Radio Bursts;
ES6: Planetary Lightning;
ES7: Joint Jupiter studies;
ES8: Cluster of galaxies \& AGNs;
and ES9:  Cluster Filament \& Cosmic Magnetism.

% Optional acknowledgements
% -------------------------
%\begin{acknowledgements}
%The standard acknowledgement, if required, is : Thank you!
%\end{acknowledgements}

%%-----------------------------
%%   Bibliography
%%-----------------------------
%%
%% The following lines are required when using BibTEX (strongly encouraged!):
\bibliographystyle{aa}  % A&A bibliography style file (aa.bst)
\bibliography{combes_S07} % your references in file: Yourfile.bib

\end{document}